\documentclass[reprint,superscriptaddress,amssymb,amsmath,aps,prl,longbibliography]{revtex4-1}

\usepackage{graphicx}
\usepackage{dcolumn}
\usepackage{bm}
\usepackage{siunitx}
\pdfoutput=1

\begin{document}


\title{Measurement of motion beyond the quantum limit by transient amplification}
\author{R. D. Delaney}
\email{robert.delaney@colorado.edu}
\affiliation{JILA, Boulder, Colorado 80309, USA}
\affiliation{Department of Physics, University of Colorado, Boulder, Colorado 80309, USA}
\author{A. P. Reed}
\affiliation{JILA, Boulder, Colorado 80309, USA}
\affiliation{Department of Physics, University of Colorado, Boulder, Colorado 80309, USA}
\affiliation{Honeywell Quantum Solutions,  Broomfield, Colorado 80021}
\author{R. W. Andrews}
\affiliation{JILA, Boulder, Colorado 80309, USA}
\affiliation{Department of Physics, University of Colorado, Boulder, Colorado 80309, USA}
\affiliation{HRL Laboratories, LLC, Malibu, California, 90265}
\author{K.~W. Lehnert}
\affiliation{JILA, Boulder, Colorado 80309, USA}
\affiliation{Department of Physics, University of Colorado, Boulder, Colorado 80309, USA}
\affiliation{National Institute of Standards and Technology, Boulder, Colorado 80309, USA}

\begin{abstract}
Through simultaneous but unequal electromechanical amplification and cooling processes, we create a method for nearly noiseless pulsed measurement of mechanical motion.  We use transient electromechanical amplification (TEA) to monitor a single motional quadrature with a total added noise $-8.5\pm2.0$~dB relative to the zero-point motion of the oscillator, or equivalently the quantum limit for simultaneous measurement of both mechanical quadratures.  We demonstrate that TEA can be used to resolve fine structure in the phase-space of a mechanical oscillator by tomographically reconstructing the density matrix of a squeezed state of motion.  Without any inference or subtraction of noise, we directly observe a squeezed variance $2.8\pm 0.3$~dB below the oscillator's zero-point motion.

\end{abstract}

\pacs{Valid PACS appear here}
\maketitle
The past ten years has seen a dramatic improvement in the ability to measure and control the quantum state of macroscopic mechanical oscillators.  Much of this progress results from advances in the parametric coupling of these oscillators to optical cavities or resonant electrical circuits. These related fields of optomechanics and electromechanics have demonstrated the ability to cool mechanical oscillators to near their motional ground state \cite{teufel2011sideband}, entangle mechanical oscillators with each other \cite{ockeloen2018stabilized, riedinger2018remote} or with other degrees of freedom \cite{palomaki2013entangling}, and create squeezed states of motion \cite{wollman2015quantum, lecocq2015quantum, Pirkkalainen2015squeezing}. To verify the successful creation of these non-classical states, electromechanical and optomechanical methods have also enabled measurements of mechanical motion with near 50\% quantum efficiency \cite{rossi2018measurement, reed2017faithful}, or equivalently an added noise equal to the zero-point motion of the oscillator, the quantum limit for simultaneous measurement of both mechanical quadratures \cite{caves1982quantum}.

These advances have encouraged notions of using non-classical states of motion to test quantum mechanics at larger scales, sensing forces with quantum enhanced precision, and enabling quantum transduction between disparate physical systems \cite{higginbotham2018harnessing}.  But as mechanical oscillators are prepared in more profoundly quantum states \cite{chu2018creation, moores2018cavity}, with finer features in oscillator phase-space, the measurement efficiency must further improve to resolve these fine features and to use them to realize a quantum advantage.  

Reaching higher levels of efficiency with existing methods is hindered by fundamental and technical limitations, which seem difficult to overcome.  In electromechanical and optomechanical devices, the state of motion can be converted without gain or added noise into a propagating electric field, and one quadrature component of the field can be measured nearly noiselessly \cite{palomaki2013entangling, rossi2018measurement}.   However, the loss experienced by the field traveling between the device and the amplifier has prevented quantum efficiency much greater than 50\%. To improve measurement efficiency, the device can be used as its own parametric amplifier, emitting an electric field that encodes an amplified copy of the mechanical oscillator's state, thereby overcoming any subsequent loss and inefficiency of the following measurement chain. Using this strategy, both quadratures can be measured simultaneously with added noise very close to the quantum limit \cite{reed2017faithful}.  For steady state monitoring of a single quadrature, backaction evading schemes are in principle, noiseless \cite{lei2016quantum, lecocq2015quantum}.  However, unwanted parametric effects, both parasitic \cite{suh2012thermally,suh2013optomechanical} and intrinsic to the electromechanical Hamiltonian \cite{supp, liao2011parametric, shomroni2018two}, have prevented measurements with noise far below the quantum-limited value.

In this Letter, we implement an efficient measurement of a single mechanical quadrature, monitoring mechanical motion with an added noise of $-8.5\pm 2.0$~dB relative to zero-point motion, and a quantum efficiency of $\eta_\textrm{q} =  88\pm 5$~\%.   By generating mechanical dynamics equivalent to the time-reverse of dissipative squeezing \cite{kronwald2013arbitrarily}, we intentionally induce mechanical instability through the electromechanical interaction, allowing for a pulsed measurement of the initial state of the mechanical oscillator.  We term this protocol transient electromechanical amplification (TEA), and demonstrate the resolution of fine features in phase space by using TEA to perform quantum state tomography \cite{lvovsky2009continuous} on a dissipatively squeezed state of the mechanical oscillator, from which we reconstruct the mechanical density matrix.  
\begin{figure}
    \centering
    \includegraphics{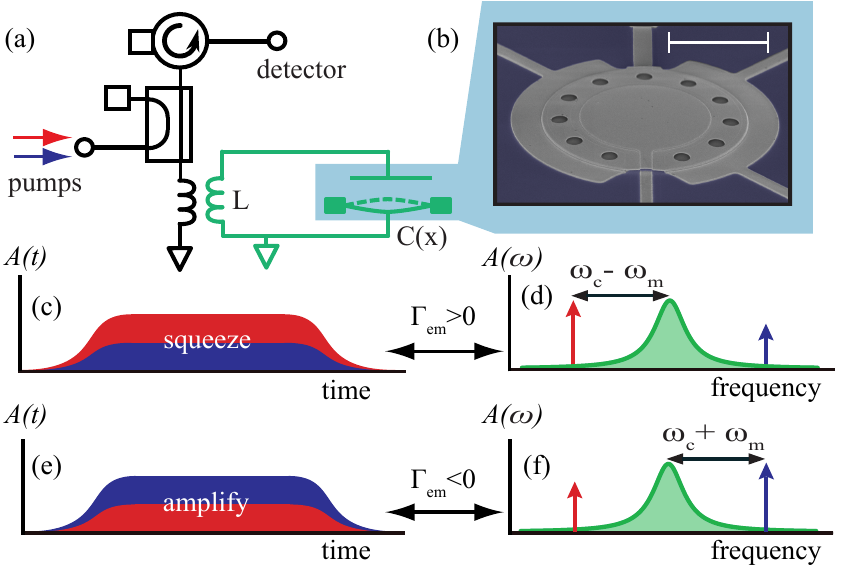}
    \caption{(a) Schematic of experiment consisting of the electromechanical circuit (green) inductively coupled to a transmission line.  Pump tones are applied through a directional coupler, while outgoing microwave signals are directed to a chain of conventional microwave amplifiers and mixer circuits, forming a microwave receiver, which adds noise much larger than zero-point fluctuations.  (b) False-color micrograph of aluminum drum.  The white bar corresponds to a distance of approximately 10 \si{\micro\meter}. (c) Time and (d) frequency domain representation of temporally overlapping dissipative squeezing pump tone amplitudes ($A(t)$ and $A(\omega)$). (e) Time and (f) frequency domain representation of transient electromechanical amplification (TEA) pump tone amplitudes.}\label{fig1}
\end{figure}

The device (shown schematically in Fig.~\ref{fig1}a) is an aluminum inductor-capacitor (LC) circuit composed of a spiral inductor and a compliant vacuum gap capacitor, which couples electrical energy to motion.  The LC circuit has a resonant frequency of $\omega_\textrm{c} \approx 2\pi\times 7.4$ GHz, and is coupled to a transmission line at a rate $\kappa_\textrm{ext} \approx 2\pi\times 3.1$~MHz.  The compliant top-plate of the capacitor (shown in Fig.~\ref{fig1}b) is free to vibrate with a fundamental mechanical resonant frequency of $\omega_\textrm{m} \approx 2\pi\times 9.4$ MHz and mechanical linewidth of $\Gamma_\textrm{m} \approx 2\pi\times 21$~Hz.  For additional device parameters and details, see the supplement \cite{supp}.  The electromechanical system is attached to the base plate of a dilution refrigerator, resulting in a mechanical occupancy of $n_\textrm{m} \leq 40$ in thermal equilibrium.  

The electromechanical circuit is in the resolved sideband regime \cite{aspelmeyer2014cavity}, enabling coherent control of motion with microwave tones.  Applying a red detuned microwave pump to the LC circuit ($\Delta \equiv \omega_\textrm{c}-\omega_\textrm{p}= -\omega_\textrm{m})$ allows for sideband cooling \cite{teufel2011sideband}, and state transfer between mechanical and microwave fields  \cite{andrews2015quantum, palomaki2013coherent}, where $\omega_\textrm{p}$ is the frequency of the pump tone.  A blue detuned microwave pump ($\Delta = +\omega_\textrm{m})$ creates  entanglement between mechanical and microwave fields \cite{palomaki2013entangling}, and realizes a quantum limited phase-insensitive amplifier of mechanical motion \cite{reed2017faithful}.  Combining these two interactions, with simultaneous application of red and blue detuned pump tones, addresses two orthogonal mechanical quadratures $X_+ = \frac{i}{\sqrt{2}}(b^\dagger - b)$ and  $X_- = \frac{1}{\sqrt{2}}(b^\dagger + b)$ independently, and enables backaction evading measurement, dissipative squeezing and TEA.
 
The type of interaction is determined by the sign of $\Gamma_{\textrm{em}}(t) = \Gamma_-(t) - \Gamma_+(t)$, where $\Gamma_\pm(t)$ are the electromechanical growth and decay rates caused by the blue (+) and red (-) detuned microwave tones respectively \cite{palomaki2013entangling}.  Dissipative squeezing occurs when $\Gamma_\textrm{em}(t)>0$, which cools the mechanical oscillator towards a squeeezed vacuum state \cite{kronwald2013arbitrarily}.  The microwave control tones  that enable dissipative squeezing are shown schematically in the time and frequency domain in Figs.~\ref{fig1}c and \ref{fig1}d.  Ideal backaction evasion occurs when  $\Gamma_\textrm{em} = 0$, where perfect constructive interference between sidebands decouples one mechanical quadrature from microwave vacuum fluctuations, producing a noiseless representation of a single mechanical quadrature in a single microwave quadrature \cite{clerk2008back}.  Finally, TEA occurs when $\Gamma_\textrm{em}(t)<0$, amplifying motion with energy gain $G\approx e^{\left| \Gamma_{\textrm{em}} \right|t}$.  Figures \ref{fig1}e and ~\ref{fig1}f show the microwave pump tones used in the time and frequency domain for TEA.  

For both TEA and backaction evading measurement, the motion of a single mechanical quadrature $X$ is encoded in a single microwave quadrature $U$.  The variance of $U$ can then be written as the sum of the noise contributions from the signal and added noise:
\begin{equation}
    \langle \Delta U^2 \rangle = G_\textrm{tot}\left(\langle \Delta X^2\rangle+\langle \Delta X_{\textrm{add}}^2\rangle \right),
\end{equation}
where $G_\textrm{tot}$ is the total gain of the microwave receiver chain in units of $V^2/\textrm{quanta}$.  If the total added measurement noise $\langle\Delta X_{\textrm{add}}^2\rangle$ is known, then the variance of the mechanical state $\langle \Delta X^2\rangle$ can be inferred.  Equivalently, by preparing a mechanical state with known variance the added measurement noise can be characterized.  For an ideal single quadrature measurement $\langle\Delta X_{\textrm{add}}^2\rangle= 0$ and $U$ faithfully records one quadrature of of the mechanical state.  Approaching this ideal behavior is highly desirable for characterizing quantum states of motion, as the number of repeated measurements required to reconstruct a quantum state grows rapidly with added noise. Furthermore, assigning meaningful uncertainties to the extracted density matrix after any inference or deconvolution procedure is complicated and subtle, diminishing confidence in the inferred state.  

 For the two special quadratures $X_\pm$, the noise properties of TEA are determined by the relative strength of $\Gamma_+$ and $\Gamma_-$.  Assuming optimal detuning of the microwave tones by exactly $\pm \omega_\textrm{m}$ and $\Gamma_\pm\ll \kappa/2$ (avoiding the strong coupling regime), the added noise $\langle \Delta  X_{\textrm{add,}\pm}^2\rangle$ referred to the input of TEA is given by
\begin{align}
    \langle \Delta X_{\textrm{add,}\pm}^2\rangle \label{eq2}\approx \frac{(\sqrt{\Gamma_+}\pm\sqrt{\Gamma_-})^2 +\Gamma_\textrm{m}(2n_\textrm{m} +1)}{2|\Gamma_\textrm{em}+\Gamma_\textrm{m}|}. 
\end{align}
In analogy with the high cooperativity limit, if  ${\Gamma_\textrm{m}(2n_\textrm{m} +1)/|\Gamma_{\textrm{em}}+\Gamma_\textrm{m}| \ll 1}$,  then $\langle \Delta X_{\textrm{add,}-}^2\rangle$ will be less than zero-point motion.  In the case where $\Gamma_-=0$, equal noise will be added to both quadratures, enabling nearly quantum limited phase-insentive amplification \cite{reed2017faithful}.  However, if the pump frequencies deviate from optimal detuning, either through an initial detuning, or through pump-power induced shifts in the circuit's resonance frequency, Eq.~\ref{eq2} is not valid, and theory including pump induced mechanical and cavity frequency shifts is required \cite{supp}. Similarly, the variance of the squeezed and anti-squeezed quadratures after dissipative squeezing $ \langle \Delta  X_{\textrm{sq,}\pm}^2\rangle$ takes the same form as Eq. \ref{eq2}, but with $\Gamma_->\Gamma_+$. 
\begin{figure}
    \centering
    \includegraphics{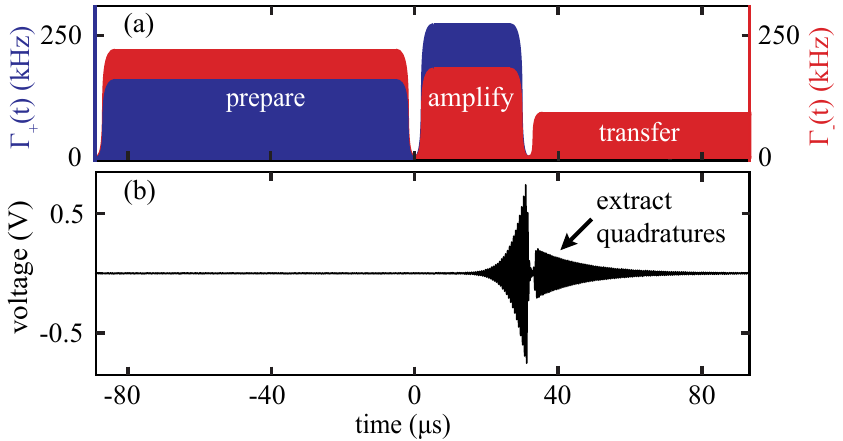}
    \caption{(a) General pulse protocol for characterization of TEA and squeezing.  The first overlapping red/blue-detuned pulses determine whether the mechanical oscillator is squeezed, cooled or allowed to thermalize with its environment.  The second set of pulses tune the gain and added noise of the measurement.  The final red-detuned pulse transfers the amplified state of the mechanical oscillator to the microwave field for quadrature extraction.  The pulse lengths are chosen so that $e^{|\Gamma_\textrm{em}|t}$ is large, and provides sufficient amplification gain or dissipative squeezing and cooling to overwhelm the thermal noise of the mechanical environment.  The pulse envelopes drawn are slower than in the experiment for visual clarity.  (b) A single experimental voltage trace of the down-converted microwave field ($\omega_\textrm{het} = 2\pi\times 1.8$~MHz) showing the resulting exponential growth and decay of the microwave field due to the amplification and transfer pulses respectively.  }\label{fig2}
\end{figure}

In Fig. 2a we demonstrate, in a three step protocol, the control of the mechanical oscillator needed to study TEA.   An initial pair of pulses prepares the mechanical oscillator in a desired state, by either sideband cooling ($\Gamma_->0$ and $\Gamma_+ = 0$), dissipatively squeezing ($\Gamma_->\Gamma_+>0$) or letting the mechanical oscillator reach equilibrium with its thermal environment ($\Gamma_+ = \Gamma_- =0$).  Following state preparation, the motion of the mechanical oscillator and the amplitude of the microwave field, are amplified by applying red and blue pumps such that $\Gamma_+ > \Gamma_-$.  After a short delay, the red-detuned pump is pulsed on to transfer the previously amplified state of the mechanical oscillator to the microwave field \cite{palomaki2013coherent}.  After further amplification by a high-electron-mobility transistor (HEMT) amplifier, and a room temperature measurement chain, the signal is mixed down to $\omega_{\textrm{het}}=2\pi\times 1.8$ MHz, allowing the two mechanical quadratures to be extracted from the exponentially decaying microwave field shown in Fig.~\ref{fig2}b.   
\begin{figure}
  \includegraphics{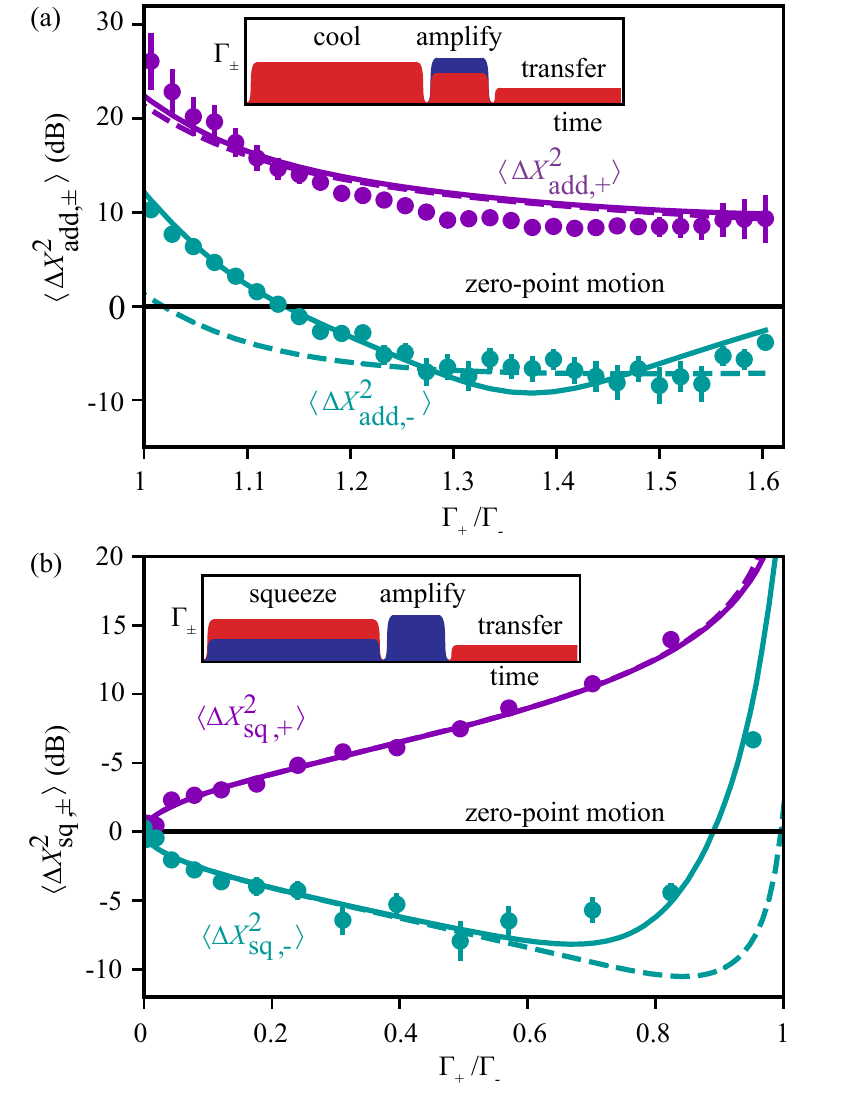}
  \caption{(a) Total added noise referred to the input of TEA relative to zero-point motion.   $\Gamma_+$ is varied, while $\Gamma_- = 2\pi\times 181$ kHz is held constant. The circles are data, while theory from Eq.~\ref{eq2} (including HEMT noise contributions) is shown without any free parameters as the dashed lines, and deviates significantly because the pump power is large enough to induce additional parametric processes.  The solid lines are theory including parametric effects (with free parameters) \cite{supp}. The inset illustrates the pulse sequence used for the inference of $\langle \Delta X_{\textrm{add,}\pm}^2 \rangle$.  Here, we obtain a minimum added noise of $\langle \Delta X_{\textrm{add,}-}^2 \rangle = -8.5\pm 2.0$~dB. (b) Inferred variance of the squeezed $\langle \Delta X^2_{\textrm{sq,}-} \rangle$ and anti-squeezed $\langle \Delta X^2_{\textrm{sq,}+} \rangle$ quadratures after dissipatively squeezing.  $\Gamma_+$ is varied, while $\Gamma_- = 2\pi \times 154$ kHz is held constant.  The minimum squeezed variance is $\langle \Delta X^2_{\textrm{sq,}-} \rangle = -7.9\pm1.4$~dB.  The circles are the data, while theory is shown without any free parameters as the dashed lines, with the expected agreement at low pump powers.  The solid lines are theory including parametric effects (with free parameters) \cite{supp}.  The inset illustrates the pulse sequence for the inference of squeezing.  }\label{fig3}
\end{figure}
  
We determine experimentally the total noise $\langle \Delta X_{\textrm{add,}\pm}^2\rangle$ added during TEA by separately preparing the mechanical oscillator in both a thermal state and through sideband cooling.  By comparing the variance of these two states in a ratio, the added noise can be inferred \cite{supp, pozar2009microwave}.  Fig.~\ref{fig3}a shows the total added noise as a function of the ratio of red and blue pump power.  With the optimal ratio of the red and blue-detuned pumps we find total added noise relative to zero-point motion of $\langle \Delta X_{\textrm{add,}-}^2\rangle =-8.5\pm2.0$~\si{\decibel}, which is equivalent to a quantum efficiency of $\eta_\textrm{q} = (1+2\langle \Delta X^2_{\textrm{add,}-}\rangle)^{-1} = 88\pm 5 \%$. We compare these results to the prediction of Eq.~\ref{eq2} with no adjustable parameters, illustrating poor quantitative agreement. We attribute this discrepancy to additional squeezing of the mechanical oscillator caused by non-linear mixing of the microwave pumps. We find good agreement in a fit to a more general theory that includes such processes \cite{shomroni2018two, supp}.  The two theories deviate significantly from each other, but TEA nevertheless achieves a minimum added noise equivalent to that predicted by the ideal case in Eq.~\ref{eq2}.  We emphasize that $\langle\Delta X^2_{\textrm{add,}\pm}\rangle$ is the total noise added by the entire measurement chain, and for $\Gamma_+/\Gamma_->1.3$ TEA has large enough gain to overwhelm the noise added by the HEMT amplifier \cite{supp}.  

Avoiding the noise associated with the simultaneous measurement of non-commuting observables is of particular importance when measuring  mechanical states with a width in phase space less than the zero-point motion of the oscillator \cite{yurke1986generating}, and is desirable for many quantum state tomography protocols \cite{lvovsky2001quantum}.  Thus, to test the effectiveness of TEA on states with variance below zero-point fluctuations, we prepare squeezed states of motion using the dissipative procedure illustrated in the inset of Fig. 3b.  To infer the total amount of squeezing, the motion is first squeezed for $90$~\si{\mu\second}, then a $30$ \si{\mu\second} blue-detuned microwave pulse ($\Gamma_+ =2\pi \times 73 $ kHz and $\Gamma_-=0$) is applied to amplify both motional quadratures.  The variance associated with zero-point motion, which must be added by the phase-insensitive amplifier, is subtracted to infer the variance of the squeezed and anti-squeezed quadratures, which is shown in Fig.~\ref{fig3}b.  We obtain a maximum inferred vacuum squeezing of $\langle \Delta  X_{\textrm{sq,}-}^2\rangle = 7.9\pm1.4$ \si{\decibel} below the zero-point motion of the mechanical oscillator.  We are able to far surpass the so-called steady state 3~dB squeezing limit both because we are using pulsed operations, and more than a single mode is involved during dissipative squeezing \cite{clerk2010introduction}.  Theory without any free parameters is plotted as the dashed lines in Fig.~\ref{fig3}b, which agrees well at low pump powers.  The solid lines show predicted squeezing when including additional parametric effects induced by nonlinear mixing of the two microwave pumps (with free parameters)  \cite{supp}.  
\begin{figure}
  \includegraphics{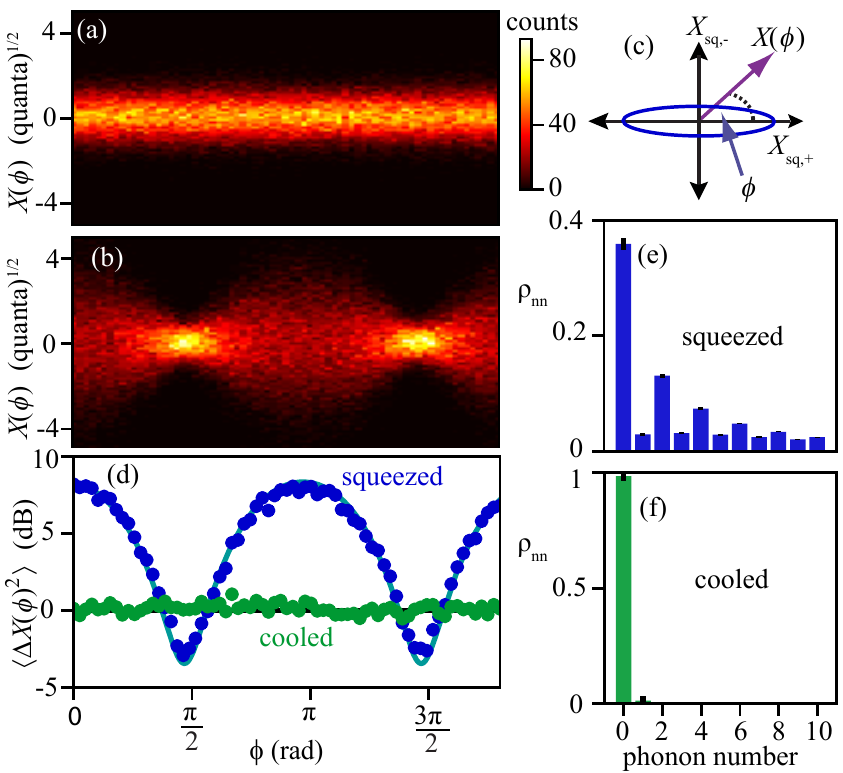}
  \caption{Measurement of mechanical squeezed and sideband cooled states with single quadrature TEA.  (a) A density plot of the marginal distributions of a sideband cooled state as a function of tomography angle $\phi$.   (b) A density plot of the marginal distributions of a squeezed state using single quadrature TEA as a function of $\phi$.  (c) The schematic shows the rotation of the single quadrature measurement axis $X(\phi)$ relative to the prepared squeezed state.  The squeezed variance is represented as the blue ellipse. (d) The total measured variance of a sideband cooled state with $n_\textrm{sb} \approx 0.02$ (green) and squeezed vacuum (blue), which exhibits squeezing $2.8\pm0.3$~dB below zero-point motion.  The data points are the circles, while theory (with no free parameters) is the solid line.  (e) Squeezed and (f) sideband cooled diagonal density matrix elements are inferred from tomographic reconstruction of the covariance matrix.  The errorbars represent 90\% confidence intervals estimated with an empirical bootstrap of the tomography data.}\label{fig4}
\end{figure}

Having demonstrated that we can prepare a squeezed state with variance below zero-point motion, the ability of TEA to resolve fine phase space features can be tested by performing quantum state tomography on the squeezed mechanical state.  By rotating a noiseless single quadrature measurement through all possible measurement axes, a set of phase space marginals can be recorded, and the density matrix can be reconstructed via quantum state tomography \cite{smithey1993measurement, vogel1989determination, hradil1997quantum, mallet2011quantum, christandl2012reliable}.  Figures~\ref{fig4}a and \ref{fig4}b show histograms of a sideband cooled ($n_\textrm{sb}\approx 0.02$) and a dissipatively squeezed state of the mechanical oscillator as a function of the tomography angle $\phi$. Figure~\ref{fig4}c demonstrates the rotation of the single quadrature measurement axis relative to the prepared squeezed state by $\phi$. 

The minimum width that can be resolved in the tomography data $\langle \Delta X_\textrm{min}(\phi)^2 \rangle$ is an important figure of merit for single quadrature measurements in the quantum regime.  In Fig.~\ref{fig4}d the total variance as a function of tomography angle is computed with theory (using independently measured parameters) shown as the solid blue line.  The squeezed quadrature has a total variance of ${\langle \Delta X_\textrm{min}(\phi)^2 \rangle=\langle \Delta X_{\textrm{sq,}-}^2 \rangle+\langle \Delta X_{\textrm{add,}-}^2 \rangle= 2.8 \pm 0.3}$~\si{\decibel} below the zero-point motion of the mechanical oscillator.  We emphasize that this represents the total reduction in noise that is present at the end of our conventional microwave receiver and no noise is subtracted to find this result.

The marginal distributions ($1.4 \times 10^5$ points in total) can be used to reconstruct the density matrix of the quantum state in the number basis.  For a general quantum state, iterative methods of tomographic reconstruction \cite{lvovsky2009continuous}--based upon maximum likelihood--are a reliable method of estimating quantum states \cite{lvovsky2001quantum}, and are guaranteed to produce a physical density matrix. However, tomographic reconstruction of squeezed states in the Fock basis requires estimating density matrix elements up to very high phonon number \cite{kim1989photon, mallet2011quantum}.  To avoid calculating large density matrices we assume Gaussian Wigner quasiprobability distributions \cite{walls2007quantum}, and estimate the density matrix through reconstruction of the covariance matrix \cite{vrehavcek2009effective}.  The covariance matrix is then used to infer the Fock basis density matrix of the mechanical oscillator.  In Figs.~\ref{fig4}e and \ref{fig4}f we plot the inferred diagonal density matrix elements for the squeezed vacuum and sideband cooled states, with the error bars on the measurements representing 90~\% confidence intervals from an empirical bootstrap procedure \cite{davison1997bootstrap, supp}.  From the density matrix we also infer the purity of the squeezed state to be $\mu\equiv1/(1+2n_\textrm{sq}) = 0.53 \pm 0.03$ \cite{supp}, where $n_\textrm{sq}$ is the equivalent thermal occupation of the squeezed state.  This demonstrates the direct resolution of features in phase space with a width approximately half that of zero-point fluctuations and the ability to resolve the squeezed character in the number basis.

Mechanical devices are increasingly being integrated into circuit QED systems as resource efficient elements, transducers and quantum memories, which offer access to new regimes of circuit QED \cite{moores2018cavity, chu2018creation}.  By directly using mechanical instability as a probe, TEA can efficiently measure motion in the presence of additional nonlinear effects.  Combining TEA with already demonstrated \cite{reed2017faithful} quantum state transfer techniques provides a path towards efficient tomography of non-Gaussian states in macroscopic mechanical oscillators.  

\begin{acknowledgments}
We acknowledge funding from AFOSR MURI grant number FA9550-15-1-0015, from ARO CQTS grant number 67C-1098620, and NSF under grant number PHYS 1734006.  We thank Lucas Sletten for help with the experiment.  We thank Brad Moores, John Teufel and Shlomi Kotler for many useful comments on the manuscript.   
\end{acknowledgments}

\end{document}


\title{Supplemental Information for: Measurement of motion beyond the quantum limit by transient amplification}

\maketitle

\hypersetup{linkcolor=black}
\tableofcontents
\hypersetup{linkcolor=red}

\clearpage

\section{Theory}\label{sec:eom}
\subsection{Two-tone electromechanical equations of motion}\label{sec:twoTone}
Starting with the linearized Heisenberg-Langevin equations of motion \cite{aspelmeyer2014cavity} in a frame rotating at a center frequency $
\omega_
\textrm{r}$ we find that
\begin{equation}
    \dot{d} = \left(-\frac{\kappa}{2} +i\Delta\right)d + ig_0\alpha(t) (b+b^\dagger)  + \sqrt{\kappa_\textrm{ex}}d_{\textrm{in}} + \sqrt{\kappa_0}f_{\textrm{in}}
\end{equation}
\begin{equation}
    \dot{b} = \left(-i\omega_\textrm{m} -\frac{\Gamma_\textrm{m}}{2}\right)b + i g_0 (\alpha^*(t)d + \alpha(t)d^\dagger) +\sqrt{\Gamma_\textrm{m}}b_{\textrm{in}}, 
\end{equation}
while in a two tone driving scheme we have
\begin{equation}
    \alpha(t) = \alpha_+e^{-i\omega_+t}+\alpha_-e^{-i\omega_-t}
\end{equation}
where $\Delta = \omega_\textrm{r}-\omega_\textrm{c}$ and $\omega_\pm = \omega_\textrm{r} \pm (\omega_\textrm{m}+\delta_\textrm{m})$.  Going to a rotating frame for the mechanical mode such that $b\rightarrow be^{-i(\omega_\textrm{m}+\delta_\textrm{m})t}$ and ignoring all fast counter rotating terms:
\begin{equation}
    \dot{d} = \left(-\frac{\kappa}{2} +i\Delta\right)d + ig_0 (\alpha_- b+\alpha_+ b^\dagger)  + \sqrt{\kappa_\textrm{ex}}d_{\textrm{in}} + \sqrt{\kappa_0}f_{\textrm{in}}
\end{equation}
\begin{equation}
    \dot{b} =  \left(i\delta_\textrm{m}-\frac{\Gamma_\textrm{m}}{2}\right)b + i g_0 (\alpha_-^*d + \alpha_+d^\dagger) +\sqrt{\Gamma_\textrm{m}}b_{\textrm{in}}.  
\end{equation}
We define the mechanical and cavity quadratures: 
\begin{equation}
 X_- = X_1 = \frac{1}{\sqrt{2}}(b^\dagger + b) 
\end{equation}
\begin{equation}
    X_+ = X_2 =\frac{i}{\sqrt{2}}(b^\dagger-b)
\end{equation}
\begin{equation}
    U_1 = \frac{1}{\sqrt{2}}(d^\dagger + d) 
\end{equation}
\begin{equation}
    U_2=\frac{i}{\sqrt{2}}(d^\dagger-d).
\end{equation}
Specializing to the case $\delta_\textrm{m}=0$ and $\omega_\pm = \omega_\textrm{c} \pm \omega_\textrm{m}$ (as in \cite{kronwald2013arbitrarily}) the linearized Heisenberg-Langevin equations can be calculated in the rotating wave approximation:  
\begin{equation}
    \dot{X}_1 = \frac{\sqrt{\kappa}}{2}\left(\sqrt{\Gamma_+} - \sqrt{\Gamma_-}\right )U_2 -\frac{\Gamma_\textrm{m}}{2}X_1 +\sqrt{\Gamma_\textrm{m}}X_{1,\textrm{in}}
\end{equation}
\begin{equation}
    \dot{X}_2 = \frac{\sqrt{\kappa}}{2}\left(\sqrt{\Gamma_+} + \sqrt{\Gamma_-}\right )U_1 -\frac{\Gamma_\textrm{m}}{2}X_2 +\sqrt{\Gamma_\textrm{m}}X_{2,\textrm{in}}
\end{equation}

\begin{equation}
    \dot{U}_1 = \frac{\sqrt{\kappa}}{2}\left(\sqrt{\Gamma_+} - \sqrt{\Gamma_-}\right )X_2 -\frac{\kappa}{2}U_1 +\sqrt{\kappa_{\textrm{ext}}}U_{1,\textrm{in}} +\sqrt{\kappa_0}f_{1,\textrm{in}}
\end{equation}

\begin{equation}
    \dot{U}_2 = \frac{\sqrt{\kappa}}{2}\left(\sqrt{\Gamma_+} + \sqrt{\Gamma_-}\right )X_1 -\frac{\kappa}{2}U_2 +\sqrt{\kappa_\textrm{ext}}U_{2,\textrm{in}} +\sqrt{\kappa_0}f_{2,\textrm{in}}.
\end{equation}
Where $\Gamma_\pm = 4g_0^2n_\pm/\kappa$ depends upon the average number of photons $n_\pm$ circulating in the LC circuit from the red (-) and blue (+) detuned pumps respectively.  Throughout this experiment we stay out of the strong coupling regime such that $\kappa/2 >> \Gamma_\pm$.  The cavity decay rate is much larger than all other decays rates in our system, thus the cavity follows the state of the mechanical oscillator, allowing for adiabatic elimination of the cavity amplitude such that $\dot{U}_1 = 0$ and $\dot{U}_2 = 0$.  This allows for simplification of the system of equations above into two independent equations for the mechanical quadratures:

\begin{equation}
    \dot{X}_1 = \frac{\Gamma_+-\Gamma_--\Gamma_\textrm{m}}{2}X_1+\sqrt{\Gamma_\textrm{m}}X_{1,\textrm{in}} +  \frac{1}{\sqrt{\kappa}}\left(\sqrt{\Gamma_+} - \sqrt{\Gamma_-}\right)\left(\sqrt{\kappa_{\textrm{ext}}} U_{2\textrm{,in}} + \sqrt{\kappa_0} f_{2\textrm{,in}}\right) 
\end{equation}
\begin{equation}
    \dot{X}_2 = \frac{\Gamma_+-\Gamma_--\Gamma_\textrm{m}}{2}X_2+\sqrt{\Gamma_\textrm{m}}X_{2\textrm{,in}} +  \frac{1}{\sqrt{\kappa}}\left(\sqrt{\Gamma_+} + \sqrt{\Gamma_-}\right)\left(\sqrt{\kappa_{\textrm{ext}}} U_{1\textrm{,in}} + \sqrt{\kappa_0} f_{1\textrm{,in}}\right).
\end{equation}

To calculate the expected variance of the mechanical quadratures the differential equations can be solved and the noise correlation functions of the cavity and mechanical quadratures can be used:
\begin{equation}
\langle f_{1,\textrm{in}}^\dagger (t) f_{1,\textrm{in}} (t') \rangle=\langle U_{1,\textrm{in}}^\dagger (t) U_{1,\textrm{in}} (t') \rangle  = \left(n_\textrm{c}  +\frac{1}{2}\right)\delta (t-t')
\end{equation}
\begin{equation}
\langle f_{2,\textrm{in}}^\dagger (t) f_{2,\textrm{in}} (t') \rangle=\langle U_{2,\textrm{in}}^\dagger (t) U_{2,\textrm{in}} (t') \rangle  = \left(n_\textrm{c}  +\frac{1}{2}\right)\delta (t-t')
\end{equation}
\begin{equation}
\langle X_{1,\textrm{in}}^\dagger (t) X_{1,\textrm{in}} (t') \rangle  = \left(n_\textrm{m}  +\frac{1}{2}\right)\delta (t-t')
\end{equation}
\begin{equation}
\langle X_{2,\textrm{in}}^\dagger (t) X_{2,\textrm{in}} (t') \rangle  = \left(n_\textrm{m}  +\frac{1}{2}\right)\delta (t-t').
\end{equation}
Computing the variance gives:
\begin{equation}
\langle \Delta X_1^2 \rangle = \langle \Delta X_1(0)^2 \rangle e^{(\Gamma_{+}-\Gamma_--\Gamma_\textrm{m})t} +     \frac{1}{2(\Gamma_{+}-\Gamma_--\Gamma_\textrm{m})}\left( (\sqrt{\Gamma_+}-\sqrt{\Gamma_-})^2(2n_\textrm{c}+1) +\Gamma_\textrm{m} (2n_\textrm{m} +1) \right)(e^{(\Gamma_{+}-\Gamma_--\Gamma_\textrm{m})t}-1)
\end{equation}

\begin{equation}
\langle \Delta X_2^2 \rangle = \langle \Delta X_2(0)^2 \rangle e^{(\Gamma_{+}-\Gamma_--\Gamma_\textrm{m})t} +     \frac{1}{2(\Gamma_{+}-\Gamma_--\Gamma_\textrm{m})}\left( (\sqrt{\Gamma_+}+\sqrt{\Gamma_-})^2(2n_\textrm{c}+1) +\Gamma_\textrm{m} (2n_\textrm{m} +1) \right)(e^{(\Gamma_{+}-\Gamma_--\Gamma_\textrm{m})t}-1).
\end{equation}

In the large gain limit, such that the energy gain $G=e^{(\Gamma_+-\Gamma_--\Gamma_\textrm{m})t}\gg1$ we find noise added at the input of transient electromechanical amplification (TEA) for the two preferred quadratures becomes:
\begin{equation}
    \langle \Delta X_{\textrm{add,}\pm}^2\rangle = \frac{1}{2|\Gamma_+-\Gamma_--\Gamma_\textrm{m}|}\left((\sqrt{\Gamma_+}\pm\sqrt{\Gamma_-})^2(2n_\textrm{c}+1) +\Gamma_\textrm{m}(2n_\textrm{m}+1)\right).  \label{eq2}
\end{equation}

If instead, we squeeze the motion of the mechanical oscillator with $e^{(\Gamma_+-\Gamma_--\Gamma_\textrm{m})t}\approx 0$, then the variance of each preferred quadrature is given by:

\begin{equation}
        \langle \Delta  X_{\textrm{sq,}\pm}^2\rangle= \frac{1}{2|\Gamma_+-\Gamma_--\Gamma_\textrm{m}|}\left((\sqrt{\Gamma_+}\pm\sqrt{\Gamma_-})^2(2n_\textrm{c}+1) +\Gamma_\textrm{m}(2n_\textrm{m} +1)\right).\label{eq3} 
\end{equation}

\subsection{Additional single mode squeezing due to detuning of microwave pumps} \label{sec:goodtheory}
The equations of motion in Section~\ref{sec:eom} are valid for $\kappa>>\Gamma_\pm$, but even if adiabatic elimination of the cavity mode remains valid, as the pump power is increased, both the cavity resonant frequency and the mechanical resonant frequency depend on the number of photons circulating in the LC circuit.  We are able to describe our data by adding in the effects of these pump induced frequency shifts.  The following analysis is similar to that in \cite{liao2011parametric} where single mode squeezing of the mechanical oscillator was demonstrated in the large detuning limit $\Delta>>\kappa$ where adiabatic elimination of the cavity field is also valid.  We operate in the limit such that $\kappa >> \Delta$ to demonstrate an equivalent result.  Assuming that $\frac{\kappa}{2} >> \Delta$, $\Gamma_\pm$, $\Gamma_\textrm{m}$, we can still adiabatically eliminate the cavity fluctuations:

\begin{equation} 
d \approx \frac{ig_0(\alpha_-b + \alpha_+b^\dagger) +\sqrt{\kappa}d_\textrm{in}}{\kappa/2 -i\Delta}, 
\end{equation}
where we have ignored the small internal loss of the cavity.  This gives an independent equation of motion for the mechanical field as:

\begin{equation}
    \dot{b} = \frac{\Gamma_+ - \Gamma_- -\Gamma_\textrm{m}}{2}b +i\left(\delta_{\textrm{m}}-\frac{\Delta}{\kappa}(\Gamma_++\Gamma_-)\right)b - \frac{2i\Delta}{\kappa}\sqrt{\Gamma_+\Gamma_-}b^\dagger + i\sqrt{\frac{\Gamma_-}{1+\frac{4\Delta^2}{\kappa^2}}}\left(1+\frac{2i\Delta}{\kappa}\right)d_\textrm{in} + i\sqrt{\frac{\Gamma_+}{1+\frac{4\Delta^2}{\kappa^2}}}\left(1-\frac{2i\Delta}{\kappa}\right)d_\textrm{in}^\dagger, 
\end{equation}
where we choose a global measurement phase and assume that $\alpha_\pm$ are both real.  In this form we can identify two additional terms in the equation for the cavity field that are caused by detuning both microwave tones from cavity resonance.  These two terms correspond to a mechanical frequency shift $\delta_\textrm{m,tot}$ and a single-mode squeezing term $\chi$ given by:
\begin{gather}
    \delta_\textrm{m,tot} = \delta_{\textrm{m}} - \frac{\Delta}{\kappa}(\Gamma_++\Gamma_-)\label{eq:deltam} \\
    \chi = \frac{2\Delta}{\kappa}\sqrt{\Gamma_+\Gamma_-}.
\end{gather}
We combine these effects into a coupled set of equations of motion to describe the two cavity quadratures.  
\begin{align}
 \begin{bmatrix} \dot{X}_1 \\ \dot{X}_2 \end{bmatrix}
 =
  &\begin{bmatrix}
   \frac{\Gamma_+-\Gamma_--\Gamma_\textrm{m}}{2}&
   \delta_{\textrm{m}} +\frac{\Delta}{\kappa}(\sqrt{\Gamma_+}-\sqrt{\Gamma_-})^2 \\
   -\delta_{\textrm{m}} -\frac{\Delta}{\kappa}(\sqrt{\Gamma_+}+\sqrt{\Gamma_-})^2 &
   \frac{\Gamma_{+}-\Gamma_{-}-\Gamma{\textrm{m}}}{2} 
   \end{bmatrix}
   \begin{bmatrix}
        X_1 \\ X_2
   \end{bmatrix}
   + \\
   &\begin{bmatrix}
   \sqrt{\Gamma_\textrm{m}}X_\textrm{1,in} + \frac{\left(\sqrt{\Gamma_+} - \sqrt{\Gamma_-} \right)}{\sqrt{1+\frac{4\Delta^2}{\kappa^2}}}\left(U_\textrm{2,in} + \frac{2\Delta}{\kappa}U_\textrm{1,in} \right) \\
    \sqrt{\Gamma_\textrm{m}}X_\textrm{2,in} +\frac{\left(\sqrt{\Gamma_+} + \sqrt{\Gamma_-} \right)}{\sqrt{1+\frac{4\Delta^2}{\kappa^2}}}\left(U_\textrm{1,in}-\frac{2\Delta}{\kappa}U_\textrm{2,in}\right)
   \end{bmatrix}.
   \label{eqn:fullTheory}
\end{align}
The additional single mode squeezing term and frequency shifts cause significant deviation from the equations of motion in section~\ref{sec:twoTone} as the power of the two tones is increased.   We recently became aware of work that describes this effect in both electromechanical and optomechanical systems \cite{shomroni2018two}.    

\section{Pump dependent cavity frequency shifts}
We estimate the cavity frequency shifts induced by the two microwave tones from optomechanical theory \cite{aspelmeyer2014cavity} by noting that there is a nonlinear dependence of the detuning from the cavity on microwave pump power, where $\alpha$ and $\beta$ are the coherent displacements of the cavity and mechanical oscillators respectively.  

\begin{equation}
    \alpha =  \frac{\sqrt{\kappa_\textrm{ex}}\alpha_\textrm{in}}{\kappa/2 - i\Delta_\textrm{eff}}
    \label{eq:implicitalpha}
\end{equation}
\begin{equation}
    \beta =\frac{-ig_0\left|\alpha\right|^2}{\Gamma_\textrm{m}/2 - i\omega_\textrm{m}},
    \label{eq:implicitbeta}
\end{equation}
\begin{equation}
    \Delta_\textrm{eff}=\Delta_0 +g_0(\beta + \beta^*).
\end{equation}
Solving these equations (neglecting $\Gamma_\textrm{m}$) we find that the effective detuning is
\begin{equation}\label{eq:deltaeff}
    \Delta_\textrm{eff} \approx \Delta_0 +\frac{2g_0^2}{\omega_\textrm{m}}|\alpha|^2 =  \Delta_0 +\frac{\kappa}{2\omega_\textrm{m}}(\Gamma_+ + \Gamma_-),
\end{equation}
demonstrating the implicit dependence of the cavity resonant frequency on the pump power sent to the circuit.  This also demonstrates that a detuned back-action evading measurement \cite{lei2016quantum} will cause degenerate parametric amplification of the mechanical mode. 
We use  the result from Equation \ref{eq:deltaeff} to solve the full set of equations in Equation~\ref{eqn:fullTheory}.  With free parameters $\delta_{\textrm{m}} = 300$ Hz and $\Delta_0 = -74 $ kHz we get good agreement between theory and experiment.  Given that there are two adjustable parameters to make the theory agree other parametric effects such as the ones described in \cite{suh2013optomechanical} may also contribute.  Thermal effects (such as \cite{suh2012thermally}) are unlikely because we see only a small shift in mechanical resonant frequency with temperature with $\frac{1}{2\pi}\frac{d\omega_\textrm{m}}{dT} \approx 4$ Hz/mK.    
\section{Variance normalization}
When preparing or measuring states at or near the limits imposed by quantum mechanics the relevant scale for the variance is the zero-point motion of the oscillator.  This is the appropriate scale for squeezing because vacuum squeezing reduces the variance in a single quadrature below the zero-point motion.  It is also the relevant scale for measurement noise, because a high gain phase preserving amplifier (simultaneous measurement of both quadratures) must add at least the equivalent of zero-point fluctuations at the input of the measurement \cite{caves1982quantum}.  Thus we normalize all variance measurements to zero-point motion  $\langle \Delta X_\textrm{zp}^2 \rangle = 1/2$.  Explicitly, this gives the measured values in decibels as:
\begin{equation}
\langle \Delta X_\textrm{add,db}^2 \rangle = 10 \log_{10}\left(\frac{\langle \Delta X_\textrm{add}^2 \rangle}{\langle \Delta X_\textrm{zp}^2\rangle}\right)
\end{equation}
\begin{equation}
\langle \Delta X_\textrm{sq,db}^2 \rangle = 10 \log_{10}\left(\frac{\langle \Delta X_\textrm{sq}^2 \rangle}{\langle \Delta X_\textrm{zp}^2\rangle}\right).
\end{equation}
In Fig.~4d of the main text we characterize the variance of the squeezed state $\langle X(\phi)^2 \rangle$ (measured by single quadrature TEA) as a function of phase angle $\phi$.  To see that we can directly measure a squeezed state without any inference we include measurement noise, but still normalize to zero-point motion:
\begin{equation}
\langle X(\phi)_\textrm{db}^2 \rangle = 10 \log_{10}\left(\frac{\langle \Delta X(\phi)^2 \rangle + \langle \Delta X_{\textrm{add,}-}^2 \rangle}{\langle \Delta X_\textrm{zp}^2\rangle}\right).
\end{equation}
The quantity $\langle \Delta X(\phi)^2 \rangle + \langle \Delta X_\textrm{add}^2 \rangle$ provides a direct measure of the squeezed state without any inference or noise subtraction.
\section{Mechanical occupancy, added noise and gain calibrations} \label{nsbnadd}
To characterize the noise added by TEA and the total amount of mechanical squeezing, we require states with known variance to calibrate these unknown values.  To accurately infer the variance of a prepared mechanical state $\langle \Delta X(\phi)^2 \rangle$ the added measurement noise $\langle \Delta X(\phi)_\textrm{add}^2 \rangle$ must be known.  Conversely to characterize the noise added by measurement, the variance of the prepared mechanical state must be known.  We do this by comparing the variance of a calibrated mechanical thermal state, to the state we wish to characterize.  In microwave engineering this is known as a Y-factor measurement \cite{pozar2009microwave}: 
\begin{equation}
r(\phi) = \frac{\langle \Delta X_\textrm{therm}^2 \rangle + \langle \Delta X(\phi)_\textrm{add}^2 \rangle}{\langle \Delta X(\phi)^2 \rangle + \langle \Delta X(\phi)_\textrm{add}^2 \rangle}. 
\end{equation}
The variance of the mechanical thermal state is:
\begin{equation}
\langle \Delta X_\textrm{therm}^2 \rangle = n_\textrm{m} +1/2,
\end{equation}
where $n_\textrm{m}$ is the independently calibrated thermal occupancy (described in  Section~\ref{sec:tempsweep}).   of the mechanical oscillator.  
\subsection{Calibration of sideband cooling and two-quadrature measurement noise}\label{sec:bothcalib}
In order to fully characterize the added noise of TEA we prepare the mechanical oscillator in states with two different temperatures: a thermal state and a sideband cooled state.  A thermal state can be prepared and calibrated by allowing the mechanical oscillator to reach thermal equilibrium (see Section~\ref{sec:tempsweep}).  To calibrate the variance of a sideband cooled state $\langle \Delta X^2_\textrm{sb} \rangle$, we measure both quadratures of motion simultaneously with the blue detuned pump $({\Gamma_+>0}, {\Gamma_-=0})$.  This nearly quantum limited amplifier adds a total noise to one quadrature of
\begin{equation}
\langle \Delta X_\textrm{add}^2 \rangle = n_\textrm{add} + 1/2, 
\end{equation}
where $n_\textrm{add}$ is any noise the amplifier adds above zero-point fluctuations.  While the variance of the sideband cooled state is
\begin{equation}
\langle \Delta X_\textrm{sb}^2 \rangle = n_\textrm{sb} + 1/2, 
\end{equation}
where $n_\textrm{sb}$ is the remaining thermal occupancy after sideband cooling.  The ratio of the measured thermal state variance to the sideband cooled variance is:
\begin{equation}\label{eqn:ratio}
r = \frac{\langle \Delta X_\textrm{therm}^2 \rangle + \langle \Delta X_\textrm{add}^2 \rangle}{\langle \Delta X_\textrm{sb}^2 \rangle + \langle \Delta X_\textrm{add}^2 \rangle} = \frac{n_\textrm{m} + n_\textrm{add} + 1}{n_\textrm{sb} + n_\textrm{add} + 1}. 
\end{equation}
From our measurements $r\gg1$, indicating that $n_\textrm{m}\gg n_\textrm{add}$, which allows us to neglect $n_\textrm{add}$ in the numerator, and solve the equation for $n_\textrm{sb}+ n_\textrm{add}$
\begin{equation}
    n_\textrm{sb} + n_\textrm{add} \approx \frac{1}{r}(n_\textrm{m} + 1) -1.
\end{equation}
Multiple measurements of this number yield a value:
\begin{equation} 
    n_{\textrm{sb}} + n_{\textrm{add}} = .05 \pm .03 \textrm{ quanta} \label{measuredOcc}
\end{equation}

We are unable to differentiate between $n_\textrm{add}$ and $n_\textrm{sb}$, but by comparing it to the minimum theoretically possible number we are able to demonstrate that we are nearly perfectly sideband cooling the mechanical oscillator.  To find this lower bound we use the theory of electromechanical sideband cooling \cite{aspelmeyer2014cavity}, and find (assuming negligible cavity occupancy) that in the resolved sideband regime the minimum mechanical occupancy after sideband cooling is:
\begin{equation}
    n_{\textrm{sb,min}} \approx \frac{\Gamma_\textrm{m}}{\Gamma_-+\Gamma_\textrm{m}}n_\textrm{m} + \frac{\kappa^2}{16\omega_m^2} = .012 \pm .002, 
\end{equation}
while equivalently the effective added occupancy at the input of the phase-preserving amplification pulse is given by:
\begin{equation}
    n_{\textrm{add,min}} \approx \frac{\Gamma_\textrm{m}}{\Gamma_+-\Gamma_\textrm{m}}n_\textrm{m}+\frac{\kappa^2}{16\omega_m^2} = .018 \pm .002. \label{nadd}
\end{equation}
This gives a total minimum remaining occupancy of:
\begin{equation}
    n_{\textrm{sb,min}} + n_{\textrm{eff,min}} = .030 \pm .004 \label{nsb}.
\end{equation}
Thus the total remaining occupancy after sideband cooling and measurement in Eq. \ref{measuredOcc} is nearly equivalent to the minimum possible value $n_{\textrm{sb}} + n_{\textrm{eff}} \approx n_{\textrm{sb,min}} + n_{\textrm{eff,min}}$, suggesting that we are cooling the mechanical oscillator to very close to the limit of resolved sideband cooling.
\subsection{Inference of mechanical squeezing}  \label{squeeze}
To infer mechanical squeezing and anti-squeezing we use TEA to measure both quadratures simultaneously ($\Gamma_+>0$, $\Gamma_- = 0$) and obtain histograms for three separately prepared mechanical states.  These histograms can then be used to calibrate the variance and added noise.  First we, prepare a mechanical thermal state with variance $\langle X^2_\textrm{therm}\rangle = n_\textrm{m}+1/2 = 36.5$ by allowing the mechanical oscillator to thermalize with its environment for $250$~ms $\approx 33/\Gamma_\textrm{m}$.  This is repeated 2048 times to obtain a histogram and estimate the variance of the thermal state in units of $V^2/\textrm{quanta}$.  With a 2 ms repetition time the same procedure is performed with a dissipatively squeezed state.  The increased speed of this measurement is enabled by the dissipative state preparation.  The variances of these two histograms are compared as a ratio:
\begin{equation}\label{eqn:ratiopm}
    r_{\pm} = \frac{\langle \Delta X_\textrm{therm}^2 \rangle +\langle \Delta X_\textrm{add}^2\rangle}{\langle \Delta X_{\textrm{sq,}\pm}^2 \rangle+\langle \Delta X_\textrm{add}^2\rangle}=\frac{n_{\textrm{m}} +n_\textrm{add}+1}{\langle \Delta X_{\textrm{sq,}\pm}^2 \rangle +n_\textrm{add}+1/2}.
\end{equation}
We assume that $n_\textrm{add}$ is equal to its lowest possible value in Eq.~\ref{nadd}.  This means that any deviation from the ideal perfect two-quadrature measurement will decrease the inferred squeezing.
For each measurement of $\langle \Delta X^2_{\textrm{sq,}\pm}\rangle$ we obtain a histogram of a sideband cooled state, allowing us to monitor that the added measurement noise remains consistent with the ideal case as in Section~\ref{sec:bothcalib}.
\subsection{Inference of the measurement noise} \label{addednoise}
We use a similar procedure to Section~\ref{squeeze} to infer the noise added by TEA.  We obtain histograms of 2048 points for a mechanical thermal state and a sideband cooled state.  This results in a ratio of variances of:  
\begin{equation}
r_\pm = \frac{n_\textrm{m}+1/2+\langle \Delta X_{\textrm{add,}\pm} \rangle}{n_\textrm{sb}+1/2 + \langle \Delta X_{\textrm{add,}\pm} \rangle},
\end{equation}
where $X_\pm$ are the two preferred quadratures of TEA.  At low gains the added noise of the HEMT is significant, and is included in the theory for single quadrature measurement noise.  For the inference of $\langle \Delta X_{\textrm{add,}\pm}^2\rangle$ we assume perfect sideband cooling as in Eq.~\ref{nsb} to provide a conservative estimate of the added noise.  For each variance data point we also make a completely separate 2048 point two-quadrature measurement of a thermal state and sideband cooled state to infer that $n_\textrm{sb}$, $n_\textrm{add}$ and $n_\textrm{m}$ remain constant over the course of the entire measurement.   

\subsection{Direct measurement of mechanical variance}
For a direct measurement of a mechanical state without any inference we use single quadrature amplification ($\Gamma_+ >0>\Gamma_- >0$) to acquire histograms of a thermal, sideband cooled and squeezed states of motion as in Section~\ref{squeeze} and Section~\ref{addednoise}.  We also separately use a two-quadrature measurement to obtain histograms of a mechanical thermal state and sideband cooled state to monitor that $n_\textrm{m}$, $n_\textrm{sb}$ and $n_\textrm{add}$ remain constant over the course of the entire measurement.  We again compute the ratio of the variance of a mechanical thermal and the squeezed state: 
\begin{equation}
    r(\phi) = \frac{\langle \Delta X_\textrm{therm}^2 \rangle+\langle \Delta X_{\textrm{add,}-} \rangle}{\langle \Delta X(\phi)_\textrm{sq}^2 \rangle+\langle \Delta X_{\textrm{add,}-} \rangle} \approx \frac{\langle \Delta X_\textrm{therm}^2 \rangle}{\langle \Delta X(\phi)_\textrm{sq}^2 \rangle+ \langle \Delta X_{\textrm{add,}-}^2 \rangle}.
\end{equation}
The approximation is valid because we independently verify that $\langle \Delta X_{\textrm{add,}-}^2 \rangle$ is negligible compared to the $\langle \Delta X_\textrm{therm}^2\rangle$.  We can then solve for the total variance of the squeezed state (as measured by TEA):
\begin{equation}
    \langle \Delta X(\phi)^2 \rangle = \langle \Delta X(\phi)_\textrm{sq}^2 \rangle+ \langle \Delta X_{\textrm{add,}-}^2 \rangle, 
\end{equation}
which is the plotted result in Fig.~4d of the main text. The blue points in Fig.~\ref{fig:YfactorMeasurement} show $r(\phi)$ as a function of tomography angle, while the red points are an equivalent measurement of a sideband cooled state (nearly mechanical vacuum) using single quadrature amplification.  The black points are the ratio of variance of a thermal state and a sideband cooled state obtained from a simultaneous measurement of both quadratures ($\Gamma_+ >0$ and $\Gamma_- =0$).  
\subsection{Estimating thermal occupancy of the mechanical oscillator} \label{sec:tempsweep}
We independently measure the thermal occupancy of the mechanical oscillator in two separate ways.  First, we use the same procedure as in \cite{reed2017faithful, andrews2015quantum}, where a very weak continuous red detuned pump is applied such that $\Gamma_\textrm{m} >> \Gamma_-$.  The ratio of the mechanical sideband power $P_\textrm{m}$ and the resulting power circulating in the cavity $P_\textrm{c}$ is measured.  The temperature sweeps in Fig. \ref{fig:TSweep1}a demonstrate that the mechanical oscillator is thermalized to 18 mK (the temperature of the dilution refrigerator) and allows us to extract the electromechanical coupling rate of $g_0 = 2\pi \times (287 \pm 11)$ Hz.  

Alternatively, we verify the temperature of the mechanical oscillator by using the blue detuned pump as a quantum limited amplifier  
\begin{equation}
    \langle \Delta X_{1\textrm{,meas}}^2 \rangle = e^{\Gamma_+t}\left(n_\textrm{m} + n_\textrm{add}+ 1\right).
\end{equation}
The measurement of a mechanical thermal state $n_\textrm{m}$ is repeated 2048 times at each different fridge temperature to compute the variance.  Drift in the gain is corrected at each temperature by measuring the exponential envelope of the amplification pulse to retrieve $\Gamma_+$.  Fig. \ref{fig:TSweep1}b shows the result of a temperature sweeps performed in this way.  The linear dependence of the variance on temperature demonstrates that the mechanical device is thermalized to the base temperature of the dilution refrigerator (16 mK).   
\subsection{Gain calibration}
From the calibration of the total noise, the gain of TEA can be estimated, and can then be compared with the expected results from theory.  In Fig. \ref{fig:gain} we compare this independently measured gain to the gain expected from $G_- = e^{(\Gamma_+-\Gamma_-)t}$ and find reasonable agreement.  The deviation for large $\Gamma_+/\Gamma_-$ may be due to additional parametric effects becoming relevant \cite{suh2012thermally, suh2013optomechanical}.  
\section{Quadrature extraction}
We perform heterodyne detection ($\omega_\textrm{het} = 2\pi\cdot 1.8$ MHz) on the microwave frequency signal emerging from the electromechanical circuit.  There are two temporally separated components of the downconverted microwave field.  First, is a pulse with an exponentially growing envelope that results from the amplification pulse.  The second exponentially decreasing microwave pulse encodes the amplified state of the mechanical oscillator that was transferred to the microwave field \cite{palomaki2013coherent}.  Both of these pulses encode the motion of the mechanical oscillator, but due to the additional parametric effects described in \ref{sec:goodtheory} the added noise is not exactly equivalent.  Fig.~\ref{fig:comparequads} shows a comparison of the added noise of TEA when measuring the amplified microwave field and the transferred mechanical field.  For all measurements in the main text the transferred mechanical field was used as it exhibited slightly better performance.
\section{Tracking measurement and squeezing axes}
The measurement and squeezing axes are controlled by the average phase $\phi_\textrm{avg} = (\phi_+ + \phi_-)/2$ of the red and blue detuned pumps.  We monitor this phase by mixing both pumps down with the local oscillator ($\omega_\textrm{LO}=\omega_\textrm{c} + \omega_\textrm{het}$) and then measuring the phase of the resulting down converted tones with frequencies $\Omega_\pm = \omega_\textrm{m} \pm \omega_\textrm{het}$ on a separate channel of the data acquisition card (see Fig.~\ref{schematic} for experimental schematic).  We use this independent measurement of the phase of the microwave tones to track the squeezing and measurement axes as a function of time.  Due to a small frequency difference between the local oscillator (Holzworth HS9000) and the red and blue detuned pump microwave generators (Agilent E8257D) the measurement and squeezing axes drift linearly with time.  We measure this phase drift and correct all data points by rotating them back to the initial measurement axis.  
\section{Quantum state tomography}
Squeezed and thermal states are inherently Gaussian states, so to infer the density matrix of these states we assume a Gaussian quasiprobability distribution and perform tomography as in \cite{vrehavcek2009effective}.  Using the set of tomography data points ${x_k(\phi_j)}$, we define the two matrix quantities:  
\begin{equation}
D = \sum_{k,j} \frac{S^T(\phi_j)|w_k\rangle\langle w_k | S(\phi_j)}{\langle w_k | S(\phi_j) G R^T(\phi_j) | w_k \rangle + \delta_\eta^2}
\end{equation}
\begin{equation}
R = \sum_{k,j} \frac{R^T(\phi_j)|w_k\rangle\langle w_k | S(\phi_j)}{\langle w_k | S(\phi_j) G R^T(\phi_j) | w_k \rangle + \delta_\eta^2} x_k(\phi_j)^2
\end{equation}
where $S(\phi)$ is the 2d rotation matrix, $\delta_{\eta_\textrm{q}}^2 = (1-\eta_\textrm{q})/(2\eta_\textrm{q})$ is a function of the quantum efficiency $\eta_\textrm{q}$ and G is the covariance matrix to be iterated over.  To find the covariance matrix describing the data we start with a vacuum covariance matrix and then iterate such that:
\begin{equation}
G^{(i+1)} = (D^i)^{-1} R^iG^iR^i (D^i)^{-1},
\end{equation}
from which the values of $D$, $R$ and $G$ are iteratively calculated.  As was shown in \cite{vrehavcek2009effective} this procedure will compute the covariance matrix that underlies the data. The covariance matrix for a squeezed thermal state is given by: 
\begin{equation}
G = \left(n_\textrm{sq} + \frac{1}{2}\right)
\begin{bmatrix}
   \cosh(2r) + \sinh(2r)\cos(\phi)&
    -\sinh(2r)\sin(\phi) \\
    -\sinh(2r)\sin(\phi) &
   \cosh(2r) - \sinh(2r)\cos(\phi) 
\end{bmatrix},
\end{equation}
with $n_\textrm{sq}$, $r$ and $\phi$ inferred from the tomographic reconstruction.  We can then calculate the matrix elements of a squeezed thermal state as in \cite{kim1989photon}.  The squeezing parameters inferred from tomography are displayed in Table \ref{tab:squeezed}.  The $90\%$ confidence intervals are found from a case resampling bootstrap algorithm \cite{davison1997bootstrap}.  
\section{Experimental apparatus}
\subsection{Electromechanical device}
The electromechanical device is mounted to the base plate of a dilution refrigerator and held at a temperature of $T\leq 18$~mK.  The design and operation of the electromechanical device is described in \cite{andrews2015quantum}.  Table \ref{tab:EMCparams} shows the relevant parameters of the electromechanical device.    
\subsection{Arbitrary microwave signal generation}
We generate the red and blue detuned microwave tones $\omega_\pm = \omega_\textrm{c} \pm \omega_\textrm{m}$ with two separate microwave generators (Agilent E8267D).  The microwave generators are pulsed on and off with an arbitrary waveform generator (Tektonix AWG 5014C).  To control the temporal envelope of the signals the microwave generators act as local oscillators on two separate double-balanced mixers (Marki-0626H), and the IF port of each mixer is controlled with a baseband signal from the AWG.  We use square pulses with Gaussian edges ($\sigma = 200$~ns) to avoid driving the mechanical oscillator into a coherent state.  The pulses are then sent through two cavity filters to reduce pump fluctuations at $\omega_\textrm{c}$.  After filtering, the pulses are split into three separate lines.  The first two lines are sent into the fridge to pump the electromechanical circuit and to cancel the reflected microwave pumps.  The third line is mixed back down against the local oscillator for independent measurement of the phase of the red and blue detuned microwave tones.  See Fig.~\ref{schematic} for a full diagram of the experimental schematic.   

The AWG controls the timing of all pulses, and runs at a 4 Hz repetition rate when measuring thermal states (to allow the mechanical oscillator to reach equilibrium with its environment) and a 500 Hz repetition rate when measuring squeezed and sideband cooled states.  To maintain phase coherence between all signals, the frequencies of all generators are set to integer multiples of 500 Hz.   
\section{Figures and tables}
\begin{table}[H]
\caption{Parameters of the electromechanical device.}
\begin{ruledtabular}
\begin{tabular}{c  l  l }
{\bf{Symbol}}	&	{\bf{Description}}	&	{\bf{Value and units}} \\
$\omega_{\mathrm{c}}/2\pi$	&	Circuit resonant frequency	&	7.376\,841 GHz	\\
$\kappa/2\pi$	&	Circuit decay rate	&	3.4 MHz	\\
$\kappa_{\mathrm{ext}}/2\pi$	 &	Circuit decay rate into the transmission line	& $3.1$ MHz		\\
$\omega_{\mathrm{m}}/2\pi$ & Mechanical resonant frequency &  9.3608 MHz \\  
$\Gamma_\mathrm{m}/2\pi$ & Mechanical decay rate & $21$ Hz  \\
$g_0/2\pi$	&	Electromechanical coupling	&	$287$ Hz \\	
\end{tabular}
\end{ruledtabular}
\label{tab:EMCparams}
\end{table}
\begin{table}[H]
\caption{Tomographically reconstructed squeezed state parameters}
\begin{ruledtabular}
\begin{tabular}{c  l  l }
{\bf{Symbol}}	&	{\bf{Description}}	&	{\bf{Value and units}} \\
$r$	& Squeezing parameter & $ 0.661 \pm 0.008$ \\
$n_\textrm{sq}$	&	Thermal occupancy	&	$0.44 \pm 0.05$ quanta	\\
$\phi$	 &	Squeezing angle	& $1.481 \pm 0.005$ 
\end{tabular}
\end{ruledtabular}
\label{tab:squeezed}
\end{table}
\begin{figure}
    \centering
    \includegraphics{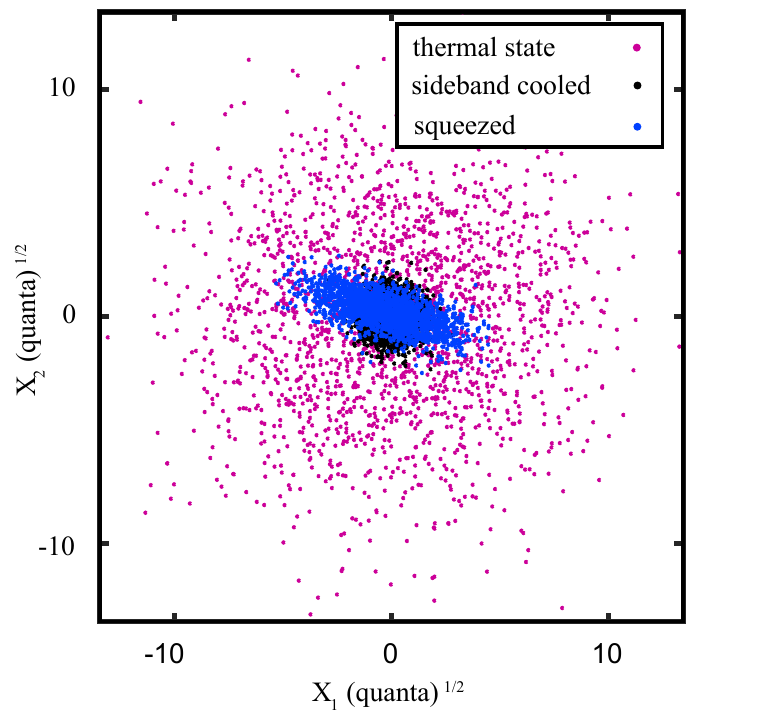}
    \caption{Typical measured histograms when inferring squezed state variance.  To estimate squeezing a thermal state (magenta points, $n_\textrm{m} = 36$) is compared with a sideband cooled state (black points, $n_\textrm{sb} > .02$) and a dissipatively squeezed state (blue).  We estimate and diagonalize the covariance matrix to retrieve the variances of the three data sets for comparison in the y-factor measurement.  This measurement was performed using only the blue detuned pump to simultaneously amplify both quadratures, thus noise equivalent to the mechanical zero-point motion is added to all of these measurements.  }\label{fig:squeezedstateHist}
\end{figure}
\begin{figure}
    \centering
    \includegraphics{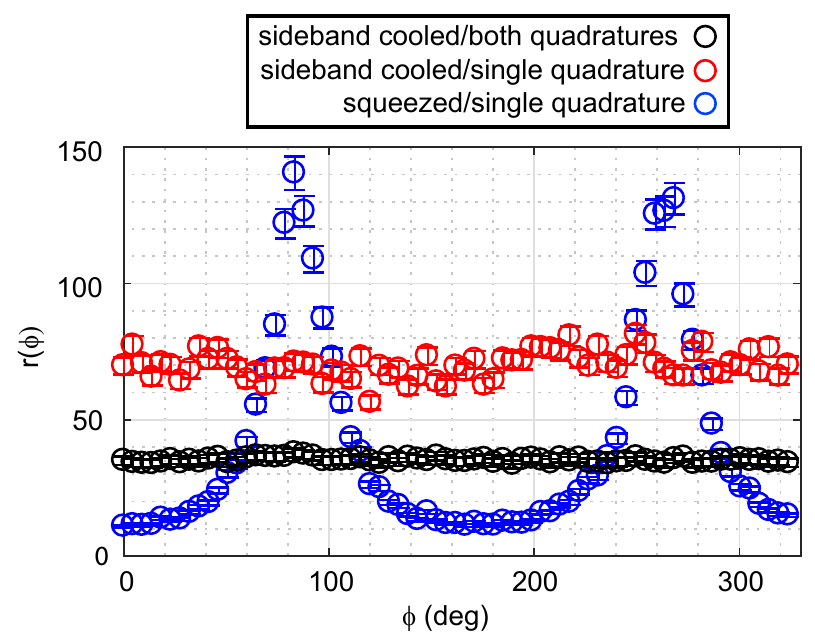}
    \caption{Comparing the variance of a mechanical thermal state ($n_\textrm{m} = 36$) to the variance of a sideband cooled state and squeezed state.  The blue points show $r(\phi)$ (see Equation~\ref{eqn:ratiopm}) for a squeezed state measured with single quadrature TEA.  The red points are $r(\phi)$ for a sideband cooled state.  The black points are $r(\phi)$ (see Equation~\ref{eqn:ratio}) for a sideband cooled state measured using two-quadrature TEA.}\label{fig:YfactorMeasurement}
\end{figure}
\begin{figure}
    \centering
    \includegraphics{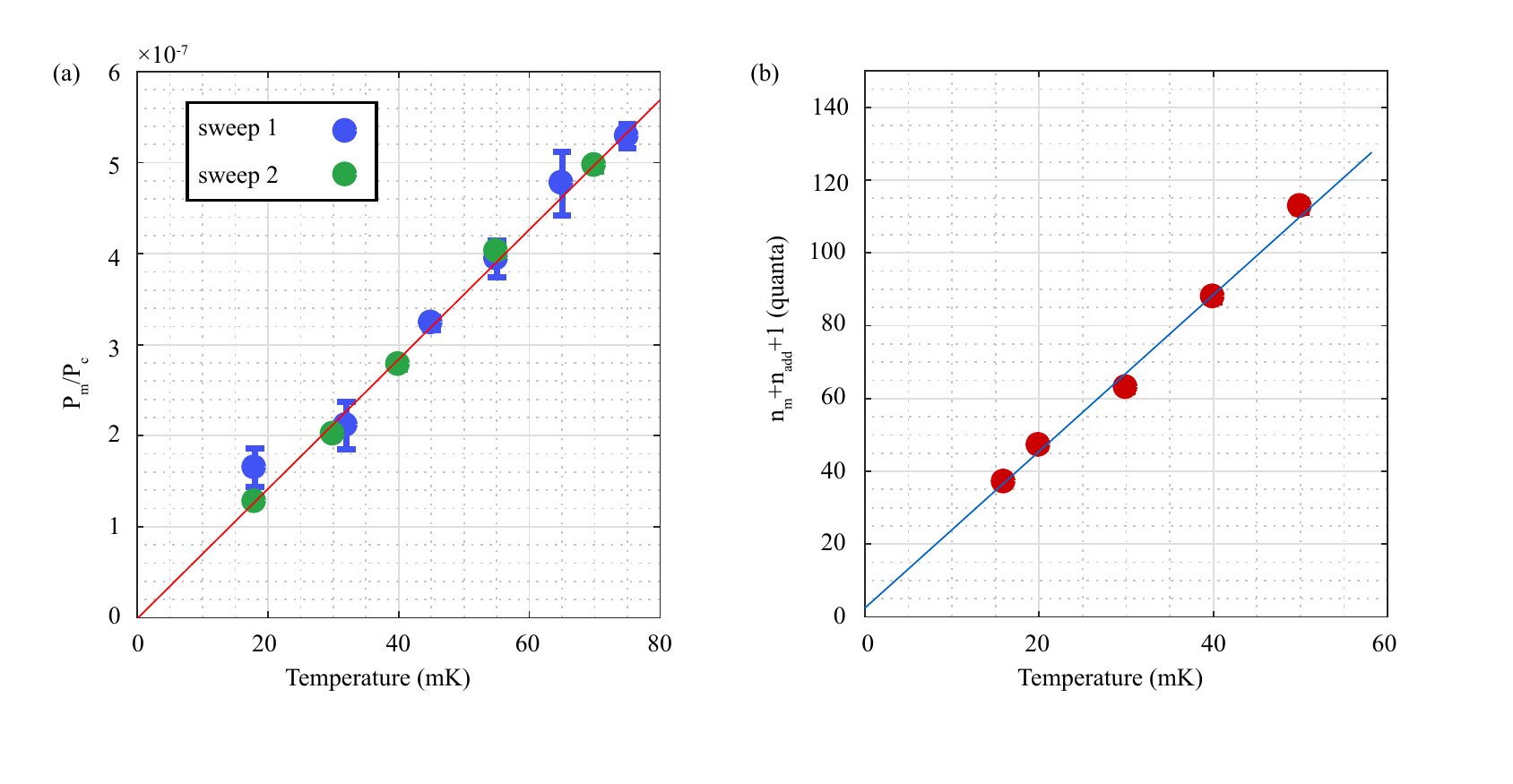}
    \caption{(a) Ratio of power in the red sideband $P_m$ to power circulating in cavity $P_\textrm{c}$ due to red-detuned pump.  This measurement is used to extract the electromechanical coupling rate $g_0 = 2\pi \times (287 \pm 11)$ and verify the steady state thermal occupancy of the mechanical oscillator. The blue and green points represent two separate temperature sweeps.  The error bars on the green points are smaller than the data points.  These measurements also demonstrate that the mechanical oscillator is thermalized to the base temperature of the fridge.  (b)  Temperature sweep using blue-detuned pump as a quantum limited amplifier, demonstrating that the mechanical oscillator remains thermalized to the base temperature of the fridge.  The linear fit extrapolated to $n_\textrm{m}=0$ yields the amplifier added noise of $n_\textrm{add} = 1.4\pm2.5$, which is consistent with the quantum limited amplification that we infer in Section~\ref{nsbnadd}.}    \label{fig:TSweep1}
\end{figure}
\begin{figure}
    \centering
    \includegraphics{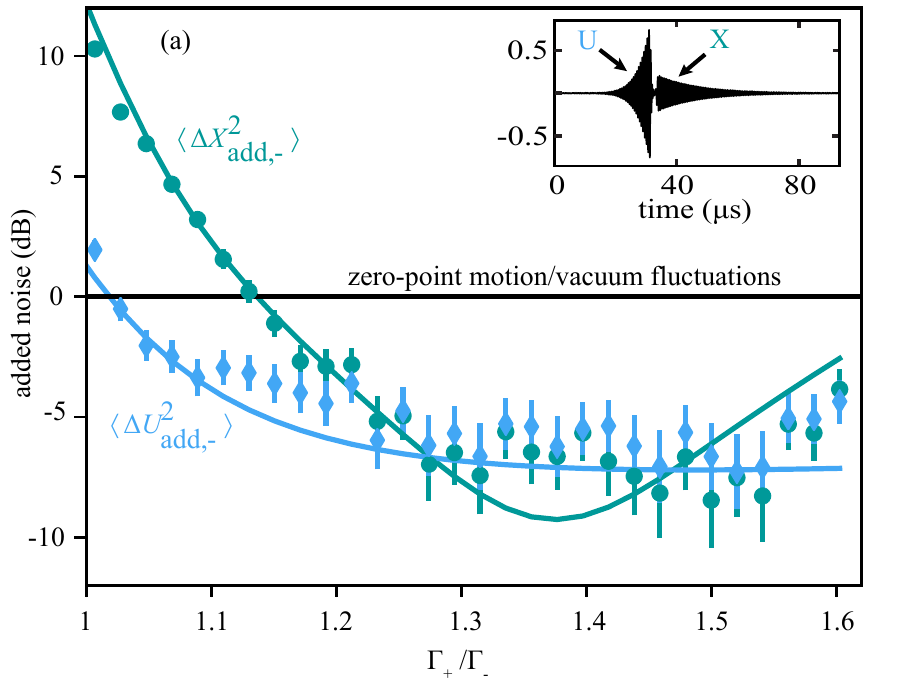}       
    \caption{Comparison of the added noise of TEA when directly measuring the microwave field $U$ or the transferred mechanical field $X$.  The blue diamonds are the added noise of TEA when using the microwave field to measure motion.  The teal circles correspond to the case where the quadratures are extracted from the transferred mechanical field. The inset demonstrates the two separate microwave pulses containing X and U.  Comparable levels of added noise are achieved, but are not equivalent due to the additional parametric effects (described in Section~\ref{sec:goodtheory}) acting on the mechanical oscillator.  }\label{fig:comparequads}
\end{figure}
\begin{figure}
    \centering
    \includegraphics{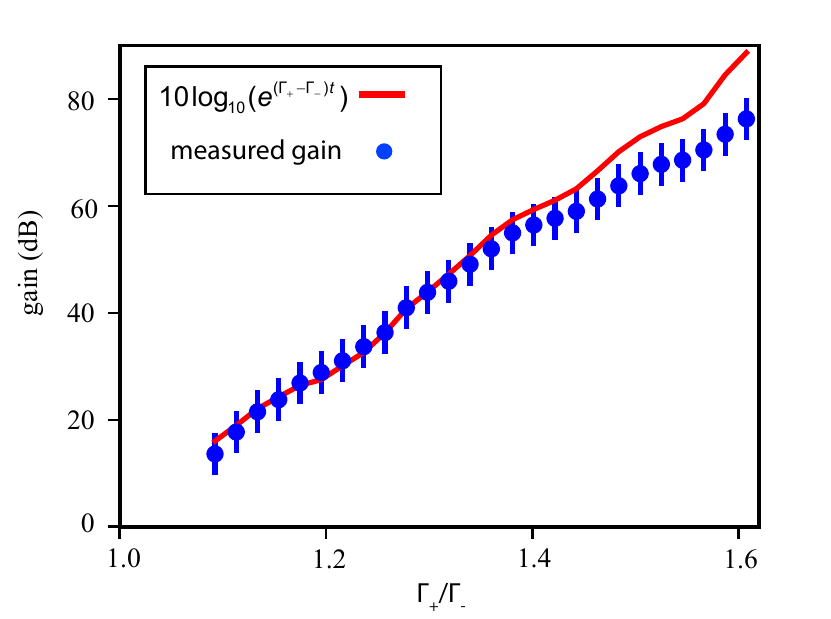} 
    \caption{Characterization of TEA gain.  The blue points are inferred from the the added noise measurements.  The red line is theory for the energy gain using only independently measured parameters.}\label{fig:gain}
\end{figure}
\begin{figure}
    \centering
    \includegraphics[width=\textwidth]{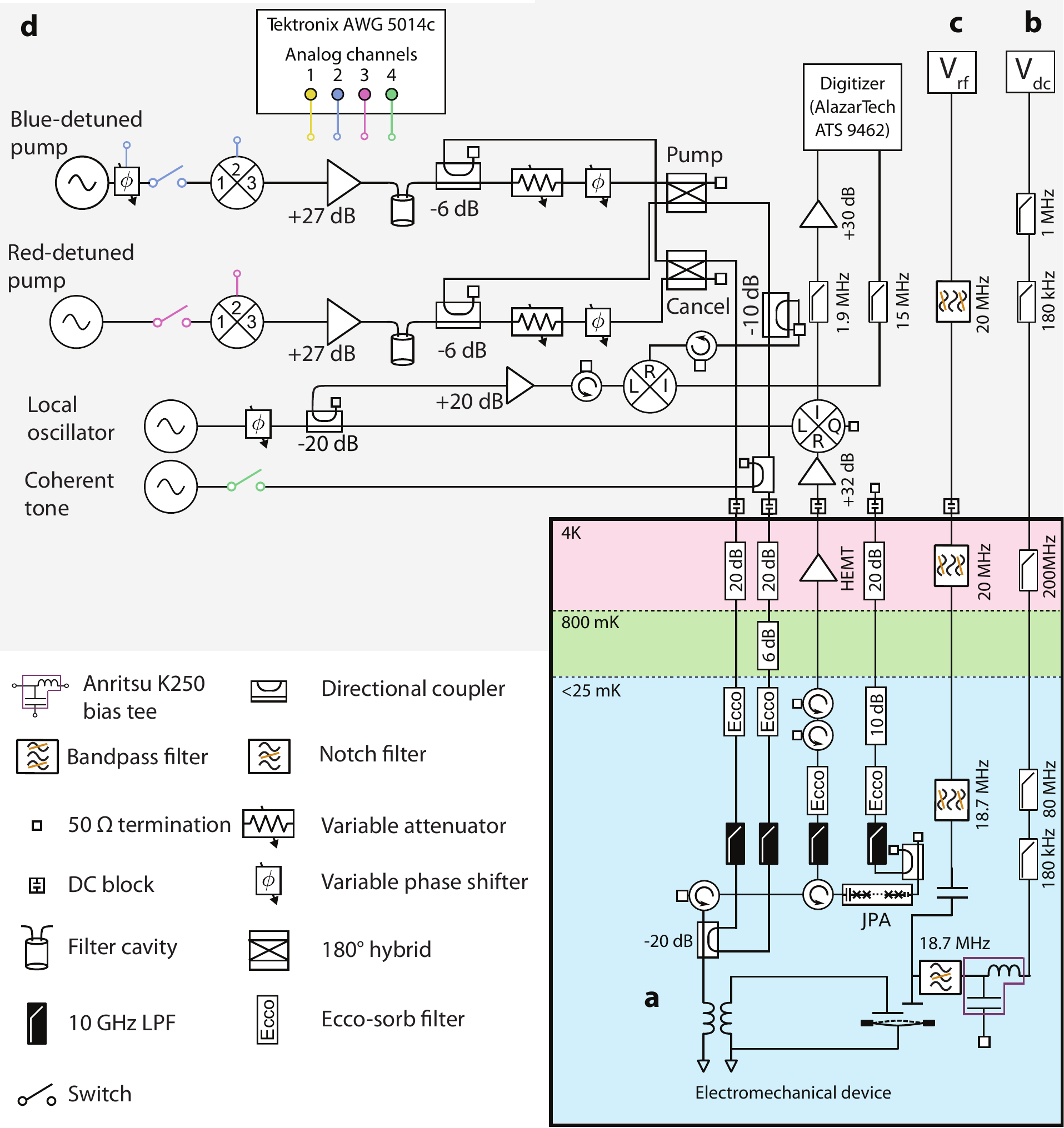}        
    \caption{Experimental schematic.  (a) Electromechanical device (b) dc actuation line that allows for the tuning of the mechanical and microwave resonant frequencies, which is not used in this work.  (c) rf actuation line allowing for rf modulation of the mechanical resonant frequency, which is not used in this work.  (d) Room temperature microwave control and measurement electronics.}\label{schematic}
\end{figure}